\documentclass[english,aps,floats,floatfix,footnotes,preprint]{revtex4} 
\usepackage[T1]{fontenc} 
\usepackage[latin9]{inputenc} 
\usepackage{amstext} 
\usepackage{graphicx} 
\usepackage{esint} 
\def\be{\begin{equation}} 
\def\ee{\end{equation}} 
 
\usepackage{babel} 
 
\begin{document}

\title{Global systematics of octupole excitations in even-even nuclei}

\author{L.M.~Robledo} 
\email{luis.robledo@uam.es} 
\homepage{http://gamma.ft.uam.es/robledo} 
\affiliation{Departamento de F\'\i sica Te\'orica, M\'odulo 15, Universidad Aut\'onoma de 
Madrid, E-28049 Madrid, Spain} 
 
\author{G.F. Bertsch} 
\affiliation{Institute for Nuclear Theory and Dept. of Physics, Box 351560, University 
of Washington, Seattle, Washington 98915, USA} 
 
\begin{abstract} 
We present a computational methodology for a  
theory of the lowest octupole excitations 
applicable to all even-even nuclei beyond the lightest. 
The theory is the well-known  
generator-coordinate extension (GCM) of the Hartree-Fock-Bogoliubov 
self-consistent mean field theory (HFB).  We  use the discrete-basis Hill-Wheeler 
method (HW) to compute the wave functions with an interaction from the Gogny 
family of Hamiltonians.  Comparing  
to the compiled experimental data on octupole excitations, we find that 
the performance of the theory depends on the deformation characteristics 
of the nucleus.  For nondeformed nuclei, the theory reproduces the energies 
to about $\pm20$ \% apart from an overall scale factor of $\approx 1.6$. 
The performance is somewhat poorer for (quadrupole) deformed nuclei,  
and for both together the dispersion of the scaled energies about the 
experimental values is about $\pm25$ \%.  This compares favorably with the 
performance of similar theories of the quadrupole excitations. Nuclei  
having static octupole  deformations in HFB form a special category.   
These nuclei have the smallest measured octupole excitation energies as 
well as the smallest predicted energies.  However, in these cases the 
energies are seriously underpredicted by the theory.  We find that 
a simple two-configuration approximation, the Minimization After 
Projection method, (MAP) is almost as accurate as the full HW treatment,  
provided 
that the octupole-deformed nuclei are omitted from the comparison. 
This 
article is accompanied by a tabulation of the predicted  
octupole excitations for 818 nuclei extending from dripline to dripline, 
computed with several variants of the Gogny interaction. 
\end{abstract} 
\maketitle 
 
\section{Introduction} 
 
The octupole excitations of nuclei have been well-studied theoretically  
on a case-by-case basis but there has never been a global study for 
a fixed Hamiltonian and well-defined computational methodology.  Such 
studies are important for several reasons.  Seeing the systematic 
trends, one can better assess the deficiencies in the Hamiltonian or 
the underlying theory, which could hopefully lead to improvements on 
both sides.  Also, the predictive power of the theory with the  
given Hamiltonians can be measured by the comparison to a large body of 
nuclear data.  In this work we carry out a study of this kind using the 
Hartree-Fock-Bogoliubov (HFB) approximation extended by Generator Coordinate  
Method (GCM).  Earlier studies of the octupole degree of freedom using  
this and similar methods are in Refs. \cite{ma83,he01,eg91,sk93b,he94,ga98}.   
A competing methodology 
is based on the quasiparticle random phase approximation; recent 
application to octupole modes may be found in Refs. \cite{se02,co03,an09}.   
For a  general review 
of the theory of octupole deformations and collective excitations, see 
Ref. \cite{bu96}.   
 
A global theory not only needs to treat the consequences of static 
octupole deformations in HFB ground states but also to treat the more 
ordinary situation where the degree of freedom appears more as a collective 
vibration of a symmetric HFB ground state.  The latter is typically treated 
by RPA or QRPA\cite{se02,co03,an09}, but the most of the studies consider a small body of 
nuclei chosen by considerations emphasizing one characteristic or another, 
for example semi-magic isotope chains.  Our study is the first to encompass 
not only magic and semimagic ordinary nuclei, but the quadrupole- 
and octupole-deformed nuclei as well.  This follows in spirit  
the studies of the nuclear quadrupole 
degrees of freedom in Refs. \cite{sa07,de10}.  We mention that our GCM coordinate is a 
one-dimensional variable labeled by the mass octupole moment.  A 
two-dimensional treatment of the octupole deformations treating the quadrupole 
deformation as a separate degree of freedom is important in theory of 
fission\cite{go05}, and is likely to play a role in spectroscopy as 
well\cite{me95}. 
 
The HFB fields and quasiparticle wave functions are assumed to have the  
following symmetries: 
time reversal, axial symmetry, and the $z$-component of isospin.  We 
can only consider even-even nuclei under these restrictions.  The 
restriction to axial symmetry is harmless in spherical nuclei, but for 
deformed nuclei it causes two problems.  The first is that theory  
only treats the $K=0$ excitations of deformed nuclei.  As we will 
see, some of the identified octupole excitations very likely have 
nonzero $K$ quantum number.  The second difficulty that arises with  
deformed nuclei is that angular momentum is not a good quantum number 
of the HFB/GCM wave function.  On a practical level, we shall compare 
the calculated excitation energies with the spectroscopic $0^+\rightarrow 
3^-$ transitions, assuming that the rotational inertias can be neglected. 
 
The calculations are carried Gogny's form of the interaction in 
the Hamiltonian.  In particular, the D1S Gogny interaction  
has been well-tested in many HFB calculations and also gives good results 
in (Q)RPA \cite{pe08} and GCM extensions of HFB \cite{ro02b}.  Specific results for 
that interaction will be presented in the text, and results for other  
Gogny interactions are provided in the 
supplementary material accompanying this article. 
 
\section{Implementing the GCM} 
 
\subsection{GCM} 
 
In the GCM, an external field is added to the Hamiltonian to generate 
a set of mean-field configurations to be taken as a basis for the 
HW minimization.  We take for the generating field the mass octupole 
operator,   
$ 
\hat Q_3 = \sqrt{\frac{4\pi}{7}} r^3 Y^3_0 (\hat r) = z^3 - 
\frac{3}{2}z(x^2+y^2). 
$.  We label the solutions of the HFB equations in the presence of the field 
$\lambda \hat Q_3$ by the expectation 
value of $\hat Q_3$,  
\be 
\langle q | \hat Q_3 | q \rangle = q. 
\ee 
For convenience, we will use the nominal value of  $\beta_3$ instead of $q$ in 
discussing the wave functions. These are related by the formula 
$q= \sqrt{9/28 \pi}(1.2)^3A^2\beta_3$.   
We also fix the (average) center-of-mass of 
the nucleus at the origin with the constraint $\langle |\hat z | \rangle = 
0$ to avoid a spurious octupole moment associated with the position of the 
nucleus. 
  
The GCM wave function is constructed by combining the 
configurations $|q\rangle$ to build a correlated wave function 
$|\sigma\rangle$.  This is expressed formally in the GCM as an 
integral over configurations 
\be 
\label{HWf} 
|\sigma\rangle = \int dq\, f_\sigma(q) | q\rangle.  
\ee  
The function $f$ in Eq. (\ref{HWf}) is to be determined by applying the variational 
principle to the expression 
\be 
E= \frac{\langle \sigma | H  | \sigma \rangle}{ \langle \sigma | \sigma 
\rangle}. 
\ee 
 
While Eq. (3) and (4) define the GCM formally, further approximations 
are required to arrive at a well-defined computational methodology. 
One way common in the literature is to keep the formal integral Eq. (3) 
and use the Gaussian overlap approximation to calculate the matrix 
elements in Eq. (4), as was done in Ref. \cite{de10} to map the 
quadrupole deformation onto a collective Hamiltonian, and in Ref. 
\cite{ro10b} for the octupole degree of freedom. A quite different way 
is the {\it discrete basis Hill-Wheeler} method, first  
carried out for the octupole excitations in Ref. \cite{ma83}.  
This method, which we will follow here, 
approximates the integral using a discrete set of 
configurations. The 
minimization is equivalent to solving the matrix eigenvalue equation 
\be 
\label{matrix-eigenvalue} 
\sum_j \langle q_{i} | H | q_{j}\rangle c_j = 
E \langle q_{i} |  q_{j}\rangle c_j. 
\ee 
The states 
will have good parity if the basis is reflection symmetric, i.e. if 
$|-q_i\rangle$ is in the basis if it contains $|q_i\rangle$. 
 
For either method one needs the overlap integrals between configurations 
$\langle q | q' \rangle$,  the matrix elements of Hamiltonian  
 $\langle q |H | q' \rangle$ and the matrix elements of one-body operators  
such as 
$\langle q | \hat Q_3 | q'\rangle$.  The basic overlap integral 
is computed with the Onishi formula\cite{onishi}.  The matrix 
elements of one-body operators and products of one-body operators 
are then evaluated using the generalized Wick's theorem\cite{balian}. 
Unfortunately, the Gogny interaction cannot be expressed in this way 
due to its $\rho^{1/3}(\vec{r})$ density dependence.   This gives rise to  
well-known ambiguities in treating the interaction as a Hamiltonian in a  
multiconfiguration space.  Of the various prescriptions  
available, we use the "mixed density" method.  Here the 
$\rho$ in the $\rho^{1/3}$ factor is replaced by $\rho_{BB}(\vec{r})$ given 
by  
\be   
\rho_{BB}(\vec{r}) = \frac{\langle q | 
\hat{\rho}(\vec{r})| q' \rangle}{\langle q | q'\rangle}  
\ee  
and the resulting 
$\vec{r}$-dependent interaction is evaluated in the usual way. 
The mixed-density prescription was  
introduced in Ref \cite{bo90} and first applied to parity-projected HFB as 
"Prescription 2" in 
Ref \cite{eg91}.   It is consistent 
with the mean field limit and is a scalar under symmetry transformations 
\cite{ro07}.  Another prescription which seems plausible at first sight 
is to use the projected density for $\rho^{1/3}$.   However, this gives 
unphysical results for octupole deformations\cite{ro10a}. 
 
While the configurations $|q\rangle$ constructed with the octupole 
constraint have 
mixed parity, the HW solutions restore the parity quantum number, 
provided that we use a basis that  
contains the both signs of $q$ in the included configurations. 
In effect, the parity projection needed to calculate spectroscopic 
properties can be obtained from the HW minimization  
without any extra 
effort.  However, as a practical matter, it is easier to define the 
parity operator in the harmonic oscillator basis and use it to  
construct $|-q\rangle$ from $|q\rangle$ thus avoiding a separate HFB  
minimization for the $-q$ configuration. 
 
The HW states of interest are  
the lowest lying even- and odd-parity states of spectrum, which we 
call $|e\rangle$ and $|o\rangle$.  Taking them to be normalized, the  
energies of ground state  $E_e$, the odd parity state $E_o$,  and the 
excitation energy difference $E_3$ are given by 
\be 
E_e = \langle e | H  | e \rangle; 
\,\,\,\, 
E_o = \langle o | H  | o \rangle; 
\,\,\,\, 
E_3=E_o-E_e 
\ee 
 
We follow the usual procedure to solve the matrix equation Eq. 
(\ref{matrix-eigenvalue}), using if necessary the  
singular value decomposition to avoid difficulties with an 
overcomplete space.   
 
One first diagonalizes the overlap matrix and transforms all of the 
matrices to the diagonalized basis.  Often there will be vectors which 
very small norms and the basis is truncated to exclude vectors whose 
norms are less than a certain value $n_{min}$.  The Hamiltonian is 
diagonalized in this basis, called the collective space, to give the HW  
energies.  The eigenvectors are used to calculate matrix elements of  
other operators between energy eigenstates. 
 
The main problem with the discrete 
Hill-Wheeler method is that the calculated values cannot be considered 
reliable unless both the range of deformations has been fully covered 
and that the singular value decomposition has been set to a robust 
truncation.  For most of the nuclei, we shall take as a  
basis the set of $\beta_3$ from -0.5 to +0.5 in steps of 0.025. 
For lighter nuclei, the range is extended from -1.2 to +1.2. 
The calculations are carried out as a function of the dimension  
$N_{basis}$ of the singular-value truncation.  There is generally 
a broad range of $N_{basis}$ for which the excitation energies have 
converged to some value; we take the value on this plateau as the 
HW result.  An example is shown in detail in the next section. 
 
The computation of the HW starting matrices is not trivial, requiring  
$N^2$ Hamiltonian matrix elements for a basis size $N$.  While this 
is not an important issue here, if one were to attempt GCM calculations 
in more than one variable, the number of states $N_{basis}$ could 
be large.  It is therefore  
of interest to investigate the accuracy of simpler approximations using 
fewer configurations. 
One of the simplest treatments is to take two configurations, $|q_e\rangle$
and $|q_o\rangle$, for the 
even-parity and odd-parity state, respectively.  The values of $q$ are 
chosen to minimize the projected energies of the configuration.  We follow 
Ref. \cite{sa07} calling this the Minimization After Projection (MAP) 
procedure.  The deformations and energies at the minima 
denoted $\beta_{3p},E_{p}$ and $\beta_{3m},E_{m}$ for the two projected
states. The MAP excitation energy is defined as
\be
E_3^{MAP} = E_m - E_p
\ee
 
One last general point of the computational procedure needs to be mentioned. 
While the individual HFB configurations are constructed with the desired 
proton and neutron particle numbers, the mixed configurations in the HW wave 
function may have slightly different expectation values of $N$ and $Z$. The 
energy depends strongly on $\langle N\rangle$ and $\langle Z \rangle$, and 
changes must be corrected for.  We do this by adding to the 
HW Hamiltonian the term $\lambda_p (\hat Z -Z) + \lambda_n (\hat N -N)$, 
where $\lambda_{p,n}$ are the nucleon chemical potentials at 
$\beta_{3p}$ \cite{bo90}. 
 
\subsection{HFB} 
 
The constrained HFB calculations were carried out using the code 
{\tt HFBaxial} written by one of us (L.M.R.).  It uses a harmonic 
oscillator basis specified by the length parameters $b_z$ and $b_t$ 
of the oscillator potential and the number of shells  
$N_{osc}$ in the basis.  For the calculations reported here 
we have taken a fixed spherical basis for all nuclei with oscillator 
length parameters 
$b_z=b_t=2.1$ fm.  The number of oscillator shells included in the 
basis is 10,12, and 14 for nuclei in the ranges $Z=[8,50],[52,82]$, and 
$[84,100]$ respectively. This is more than enough to provide converged  
results for energy differences.  We report on the results for the D1S Gogny  
interaction in sections below.  More detailed results for the D1S 
as well as for other interactions of the Gogny form are given in  
the supplemental material \cite{epaps}.

\section{Examples} 
 
In this section we will go through the details for four examples 
illustrating the application to a spherical nucleus, $^{208}$Pb,  
a well-deformed nucleus, $^{158}$Gd, the nucleus $^{226}$Ra whose HFB 
ground state has a static octupole deformation, and a light nucleus 
having a very large transitional octupole moment, $^{20}$Ne. 
A summary of the results for these nuclei is given in Table \ref{examples} 
at the end of this Section. 
 
\subsection{$^{208}$Pb} 
 
\begin{figure} 
\includegraphics [width = 9cm]{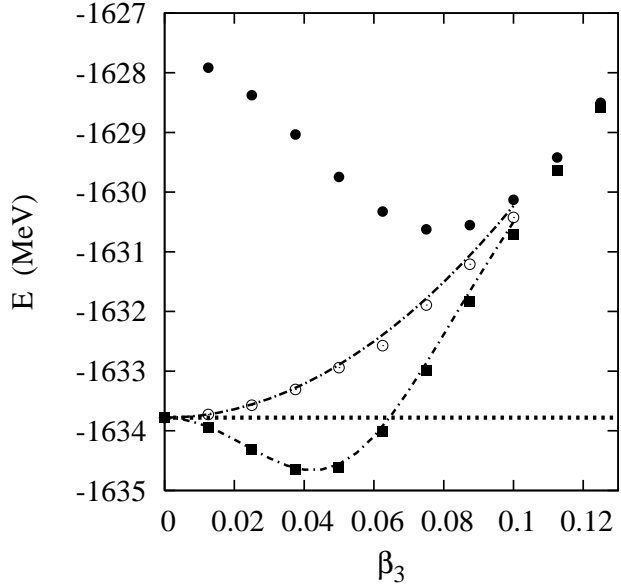} 
\caption{\label{208Pb} 
Energy of $^{208}$Pb as a function of octupole deformation $\beta_3$. Open 
circles: HFB energy of constrained configurations ;   
Solid squares: energy $E_e$ of the  
even-parity projected wave function; Solid circles:   
the odd-parity projected energy $E_o$.   See the Appendix for explanation 
of the fitted lines. } 
\end{figure}

The nucleus $^{208}$Pb is a paradigm for a doubly magic nucleus. 
It is one of the very few nuclei whose first excited state has  
$J^\pi = 3^-$ quantum numbers.  The excitation energy is 2.62 
MeV and the transition rate is strongly collective with strength of  
$B(E3,\uparrow) = 0.611 $ e$^2$b$^3$ 
or 34 
Weisskopf units\cite{wu}.  For the theory, we first shown HFB and  
projected energies of the GCM configurations in  Fig. \ref{208Pb}. 
The minimum energy projected configurations, ie. the MAP states, are at 
$\beta_{3p} \approx 0.0375$ and $\beta_{3m}\approx 
0.075$. 
One sees that the energy of the ground state is lower by projecting from 
a nonzero $\beta_3$; the associated correlation energy  
has the order of magnitude of one MeV.  The MAP approximation to the 
excitation energy $E_3$ is given by the difference of the minima of 
the plus- and minus-projected energy curves, which is about 4.2 MeV. 
 
\begin{figure} 
\includegraphics [width = 9cm]{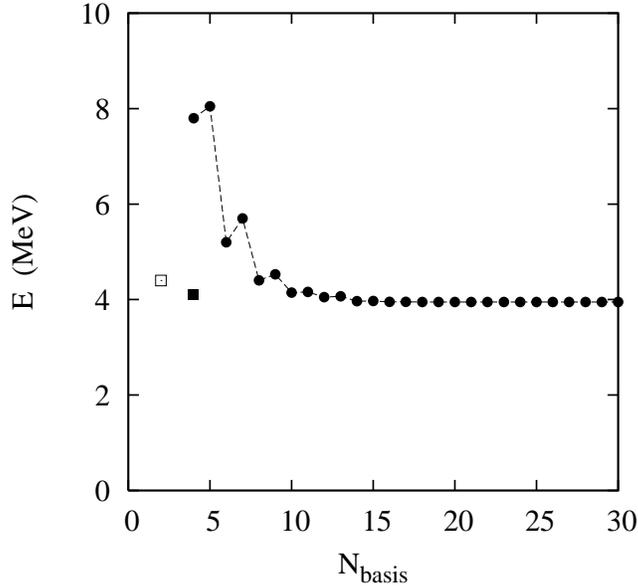} 
\caption{\label{basis} 
Excitation energy $E_3$ in $^{208}$Pb as a function of the   
configuration space choice. Solid circles: HW using the singular  
value decomposition to keep $N_{basis}$ states;  solid square: HW 
with the two MAP states; open square: Energy difference of the two 
MAP states. 
} 
\end{figure}

To see how the calculated $E_3$ depends on the basis, we show it in Fig. 
\ref{basis} as a function of $N_{basis}$.  The difference of MAP energies 
is the open square, and solid circles show the results with various 
truncations.  The full basis set is comprised of the 
41 configurations between $\beta_3=-0.5$ to 
$\beta_3=+0.5$ in steps of 0.025.  The truncation is carried out 
by the singular-value decomposition. 
One sees that the energy has converged at about $N_{basis}\approx 14$ 
and the numerics remain stable up to much larger values.   
The converged energy, 4.0 MeV, is fairly close to the difference of  
MAP energies.  In fact, one can do even better in the 4-dimensional 
space allowing the MAP configurations to mix.  This is shown as the 
solid square in the figure.  We note that our excitation energy of  
4.0 MeV is close to the value found in  
Ref. \cite{he01} using the GCM/HW method but with 
the Skyrme SLy4 interaction. 
 
\begin{figure}  
\includegraphics [width = 9cm]{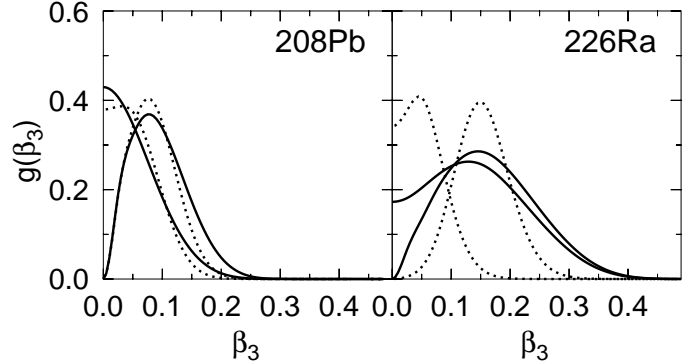}  
\caption{\label{wfn}  
Wave function amplitudes.  See text for explanation.  
}  
\end{figure}

We see here that the MAP could be a very useful simplification, but 
its validity depends on the circumstances. 
It is instructive to examine the 
GCM/HW wave function and compare it with MAP.  These are shown in 
Fig. \ref{wfn}, for both the ground state and the odd-parity excited state.   
The wave function amplitudes are formally defined by the 
integral 
\be 
g_ \sigma (\beta_3) = \int d \beta_3' {\cal N}^{1/2} (\beta_3, \beta_3') 
f_\sigma (\beta_3') 
\ee 
where $f$ is normalized $ 1= \int d\beta_3 \,d \beta_3' {\cal N} 
(\beta_3, \beta_3') 
f_\sigma (\beta_3') f_\sigma(\beta_3) $. 
The above relation establishes the connection between the standard  
GCM amplitudes $f$ with the amplitudes $g$ entering the expansion of  
the GCM wave functions in terms of orthogonal states  
$|q\rangle_\textrm{orth} = \int dq' {\cal N}^{-1/2} (q, q')  
|q'\rangle$. The square root of the norm overlap has to be understood  
in terms of the relation $\int dq'' {\cal N}^{1/2}(q,q'') {\cal  
N}^{1/2} (q'',q') = {\cal N} (q,q')$. The ground and excited state  
wave functions can be distinguished by the amplitude at $\beta_3=0$,  
which is finite for the even-parity ground state and zero for the  
odd-parity excited state. The HW wave function and the MAP  
approximation are shown as solid and dashed lines, respectively.  It  
is clear that the MAP configuration is a good approximation to the  
full wave function of both the ground and excited states, for this  
particular nucleus.  

More insight into the collective physics of the octupole degree of 
freedom can be obtained comparing with simple models of the excitation 
(See Appendix).  If the configuration energies and interactions can 
be treated as quadratic functions of the deformation coordinate, 
and the matrix elements between different configurations can be 
treated by the GOA, the GCM/HW reduces 
to the RPA and is exact.   The line through the HFB energy  
curve in Fig. \ref{208Pb} is a quadratic fit.  It appears to be well satisfied.   
Also, the energy of the even-parity projected configuration follows well 
the predicted dependence according to the GOA, Eq. (\ref{map-e}). This 
shown as the line through 
the even-parity projected energies in the figure.   
Thus two of the conditions are met to reduce the GCM/HW  
theory to an RPA of a single collective state.

\subsection{$^{158}$Gd} 
\begin{figure} 
\includegraphics [width = 9cm]{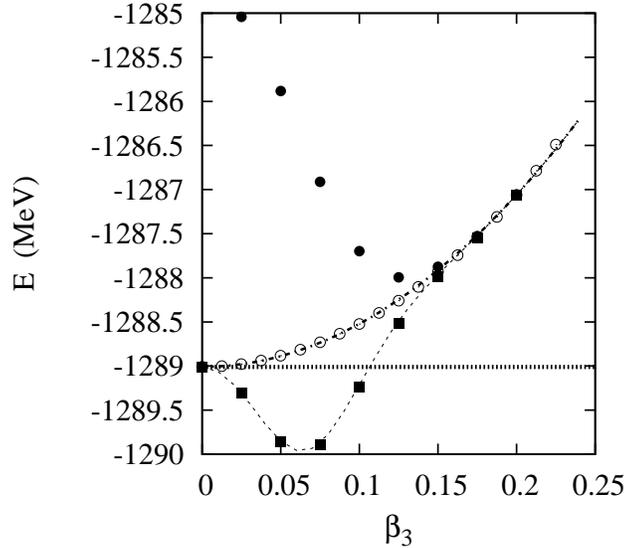} 
\caption{\label{158Gd} 
Energy of $^{158}$Gd as a function of octupole deformation $\beta_3$. Open 
circles: HFB energy of constrained configurations ;   
Solid squares: energy $E_e$ of the  
even-parity projected wave function; Solid circles:   
the odd-parity projected energy $E_o$.   The line along the HFB values 
is the function   $E_q = E_0 + K_1 \beta_3^2$ with $K_1=48.8$ MeV fitted 
to the values $\beta_3 \le 0.05$.  The line along the 
$E_e$ values is the fit motivated by the Gaussian overlap approximation, 
$ E_e = E_q - K_2 \beta^2 / (1.0 + \exp( \alpha \beta^2))$, with 
$K_2$ and $\alpha$ fitted.  
} 
\end{figure}

Our example of a strongly deformed nucleus is $^{158}$Gd.  It has 
a  $3^-$ excitation at 1.04 MeV with a transition strength 
$B(E3 \uparrow) = 0.12$ e$^2$b$^3$.  The energies from the GCM calculation 
are shown in Fig. \ref{158Gd}.  Overall, the energy curves look quite 
similar to those for $^{208}$Pb.  The HFB curve is also well 
fit by a quadratic dependence on $\beta_3$ but the curvature here is  
much shallower. 
The projected energy function  $E_e(\beta_3)$ also has a similar shape to the 
curve for $^{208}$Pb, and can be fitted by the same functional form,  
Eq. (\ref{simple-h}).  The ratio of MAP minimum points is found to be 
$\beta_{3p}/\beta_{3m}\approx 2$, similar to the situation for $^{208}$Pb. 
The excitation energy $E_3$ comes out to about 1.7 MeV, much smaller than 
the $^{208}$Pb value.  This is to be expected in view of the softer  
HFB curve.  The correlation energy of the ground state, $E_0 - E_e$, is 
similar to the $^{208}$Pb value, about one MeV.   
Experimentally, the situation is complicated by the deformation and 
the splitting of the octupole strength into different $K$-bands. 
There are three negative parity bands known experimentally at low 
energy. There is a $K=1^-$ with an $1^-$ state at 977 keV, a $K=0^-$ 
with the $1^-$ state at 1263 keV and finally a $K=2^-$ with a $2^-$ 
state at 1793 keV. Our excitation energy of 1.7 MeV should be compared 
with the 1263 keV of the $1^-$ state of the $K=0^-$ band. The 
theoretical value is stretched by a factor 1.4 with respect to the 
experimental value (see discussion below). Note that the measured 
octupole transition at 1.04 MeV is not relevant for the comparison because 
it corresponds to a different $K$ value.

\subsection{$^{226}$Ra} 
\begin{figure} 
\includegraphics [width = 9cm]{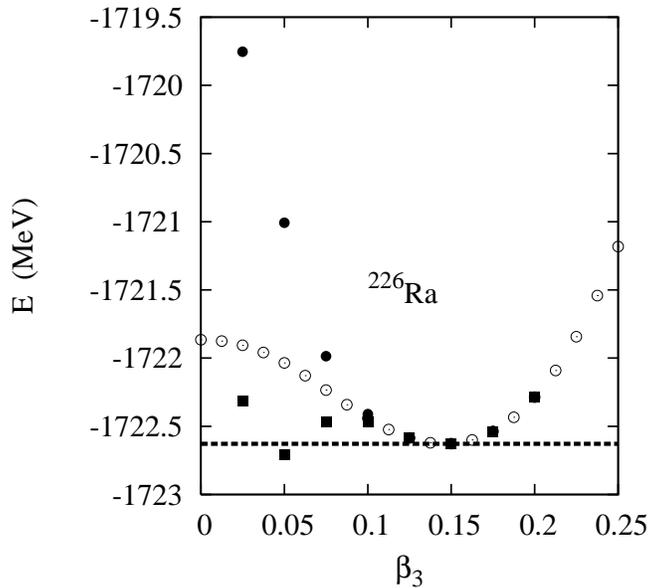} 
\caption{\label{226Ra} 
Energy of $^{226}$Ra as a function of octupole deformation $\beta_3$ 
as in Figs. \ref{208Pb},\ref{158Gd}g.  A very similar plot is shown 
in Fig. 3 of Ref.~\cite{eg91}.   
} 
\end{figure}

$^{226}$Ra has the lowest $3^-$ excitation energy of any in 
the compilation \cite{ki02}, $E_3= 320 $ keV.  It also has the highest 
transition strength in the compilation, $W(E3)= 54$ Weisskopf 
units\cite{wu}. 
On the theory side, the nucleus is predicted to deformed both in  
the quadrupole ($\beta_2\approx 0.3$) and the octupole degrees of freedom.   
The HFB/GCM 
energy curve, shown in Fig. \ref{226Ra}, has a minimum at $\beta_3\approx 0.13$.

This nucleus is very interesting for our survey, not only because of the static octupole 
deformation, but  because the theory is seen to fail badly if the 
large amplitude fluctuations are not properly accounted for. 
The predicted excitation energies for different 
treatments of the GCM configurations are shown in Table \ref{226Ra-betas}. 
The most naive theory (top line) would ignore the 
GCM construction and simply take the HFB minimum and project from that. 
The overlap $\langle -q | q\rangle$ at the HFB minimum is essentially zero 
and the $E_3$ comes out less than 1 keV.  In the next approximation 
we consider (second line), we take the single configuration that gives 
the MAP ground state.  Here the deformation is much closer to zero. 
However, the $E_3$ calculated as the difference between the 
even and odd projected states is now far too large, 1.7 MeV.  Of course 
in the full MAP approximation we should take the configurations at 
different $\beta_3$ for odd and even projections.  This is done in  
line 3 of the Table, and now the $E_3$ has the correct order of 
magnitude.  Adding more configurations, the valued do not change much 
on an absolute MeV scale, but on a relative scale there is a considerable 
change.  The most complete HW treatment, on the bottom line, underpredicts 
the energy by a factor of $\approx 2$. 
We also show the HW and MAP wave functions in Fig. \ref{wfn}.  It is 
clear that the full wave functions are far from harmonic and 
that the MAP approximation fails badly.  
 
\begin{table}[htb] 
\begin{center} 
\begin{tabular}{|c|c|lll|} 
\hline 
$N_q$ & $\beta_3$    &  $E_e$ &  $E_o$ & $E_3$  \\ 
\hline 
1   &  0.15 &  -1722.63  & -1722.63  & 0.00   \\ 
1   &  0.05 &  -1722.71 MeV & -1721.01 & 1.7 MeV \\ 
2   &  0.05,0.15 & -1723.43  &  & 0.37  \\ 
3   &  0.05,0.1,0.15  & -1723.45 &   & 0.31 \\ 
4   &  0.025,0.075,0.125,0.175 & -1723.53  &  & 0.22    \\ 
12  &  [-0.5,0.5] & & & 0.16\\ 
\hline 
\end{tabular} 
\caption{\label{226Ra-betas} Calculated energies of $^{226}$Ra with various choices of  
the configuration set. 
} 
\end{center} 
\end{table}

\begin{figure} 
\includegraphics [width = 9cm]{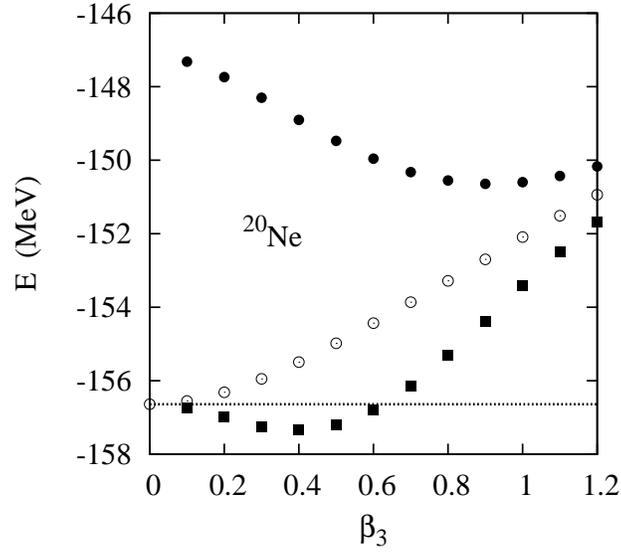} 
\caption{\label{020Ne} 
Energy of $^{20}$Ne as a function of octupole deformation $\beta_3$ 
as in Figs. \ref{208Pb},\ref{158Gd},\ref{226Ra}.  
} 
\end{figure} 
\begin{figure} 
\includegraphics [width = 9cm]{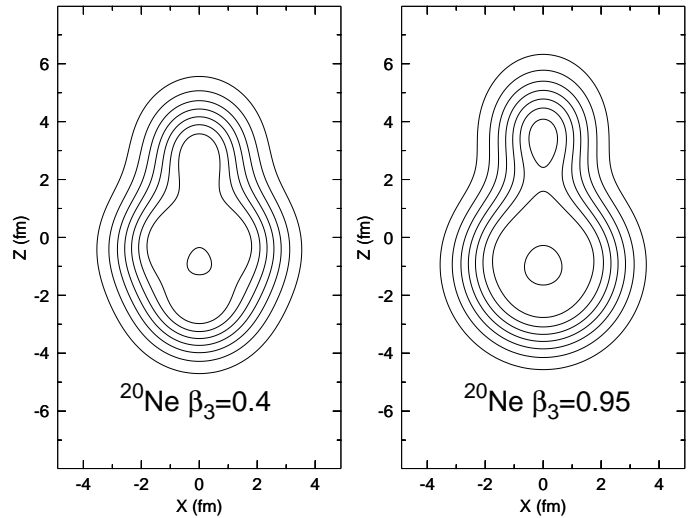} 
\caption{\label{20Ne_density} 
Nucleon density distribution in $^{20}$Ne at $\beta_{3p}$ (left) and  
$\beta_{3m}$ (right). 
} 
\end{figure} 
 
\def\ne{$^{20}$Ne} 
\subsection{$^{20}$Ne} 
\begin{figure} 
\includegraphics [width = 9cm]{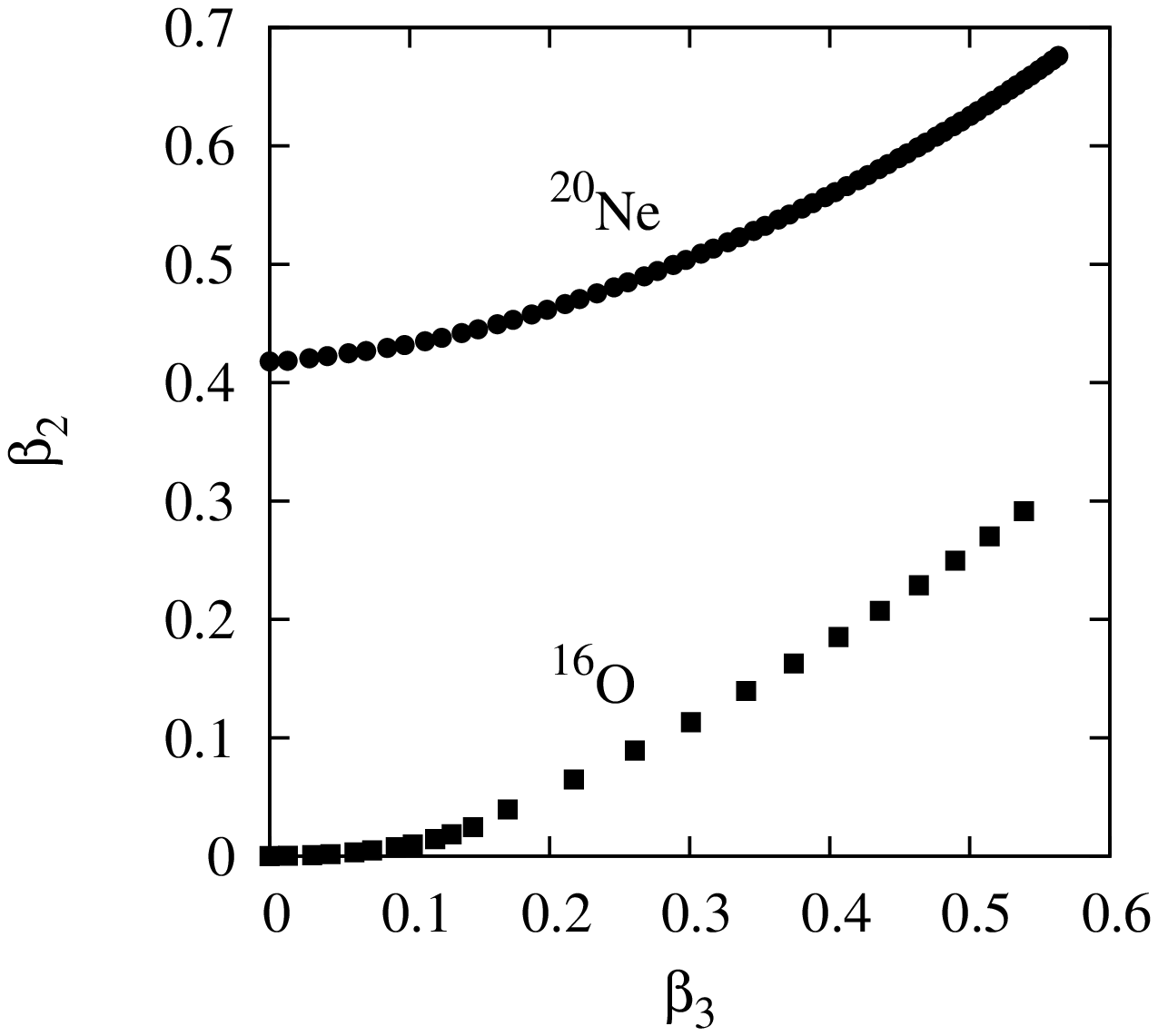} 
\caption{\label{b23} 
Deformation of the octupole-constrained HFB configurations for $^{16}$O 
and $^{20}$Ne. 
} 
\end{figure}

\ne~illustrates some differences that one sees in treating light 
nuclei by the GCM/HW, first studied by this method in Ref. \cite{ma83}. 
Due to the incipient alpha clustering, the  equilibrium 
octupole deformation of the projected configurations can be very large. 
The HFB and projected energies are 
shown in Fig. \ref{020Ne}.  Note that the HFB energy deviates from 
a quadratic dependence on the deformation, and looks almost linear at  
large $\beta_3$.  Fig. \ref{20Ne_density} shows the density distribution 
at the two projected minima.  One sees a compact localized density, 
suggestive of an alpha particle, outside a nearly spherical core. 
Since the alpha emission threshold is rather low in this nucleus, 
one should expect a softness in with respect to the generator 
coordinate corresponding to alpha cluster separation.  In a multipole 
representation, this requires changing both the quadrupole and the 
octupole deformation.  This is in fact what occurs in our GCM wave 
functions.  Fig. \ref{b23} shows their deformations in the two 
multipolarities.  The coupling of the multipolarities can cause 
problems, however.  We will come back to this in the Appendix, referring 
to the coupling in $^{16}$O, also shown on the figure. 

\begin{table}[htb] 
\begin{center} 
\begin{tabular}{|c|lll|llc|} 
\hline 
Nucleus  & \multicolumn{3}{c|}{$E_3$ (MeV)}  & 
\multicolumn{3}{c|} 
{ $W(E3)$}\\ 
 &  Exp.  &  Present & Other   & Theory & Eq. & Exp.\\ 
\hline 
$^{20}$Ne & 5.6  & 6.7 & 5.2$^{a}$ & 12. &  (\ref{be3def}) & 13.\\ 
$^{208}$Pb &2.6  & 4.0 & 4.0$^{b}$  & 53.&  (\ref{be3sph})& 34.\\ 
$^{158}$Gd & 1.04  & 1.93    &   & 11.6   & (\ref{be3def})& 12.  \\ 
$^{226}$Ra & 0.32  & 0.16 &   &43.&  (\ref{be3def}) & 54. \\ 
\hline 
\end{tabular} 
\caption{\label{examples}  
Summary of results for the four examples discussed in the text. 
References for column 4, other theory: a) \cite{ma83}; b) \cite{he01}. 
} 
\end{center} 
\end{table}

\section{Systematics} 
 
\begin{figure}   
\includegraphics [width = 9cm]{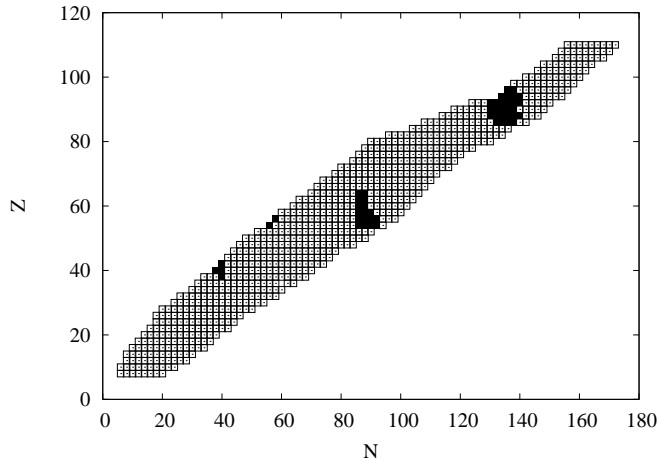}%
\caption{\label{static-b30} 
Chart of the nuclides shown those calculated in the present study. 
Those in black have static octupole  
deformations in HFB.  Except for the nuclei near $N\sim Z \sim 40$, 
the nucleon numbers correspond well to the numbers 56, 88, and 136  
listed in Ref. \cite{bu96} as especially favorable for octupole 
deformation. 
} 
\end{figure}

We have applied the HFB/GCM/HW theory across the chart of nuclides  
including 818 nuclei between $8\le Z\le 110$.  About 6\% of them 
are octupole deformed in the HFB ground state.  The nuclei are 
shown in  
Fig. \ref{static-b30}.    
Favorable conditions for static octupole deformation occur 
when a high-$j$ intruder orbital is close to an opposite-parity orbital 
with three units less of orbital angular momentum near the Fermi 
energy\cite{bu96}, which happens for $Z$ and $N$ values around 36, 56, 
88, and 134.  The regions  
around Ba and Ra are well-known in earlier studies.  We also find 
static deformations near $^{80}$Zr and 
near $Z\approx N \approx 56$ (for this region, see also Ref. \cite{he94}. 
There are also calculations in the literature reporting static octupole 
deformations in other regions as well\cite{mo08,zh10}. 
In any case,  the HFB deformation is not an observable.   
Physically, one can only measure excitation energies and transitions 
strength. These are compared with experiment in the two subsections 
following. 

\subsection{Excitation energies}  
 
\begin{figure}   
\includegraphics [width = 9cm]{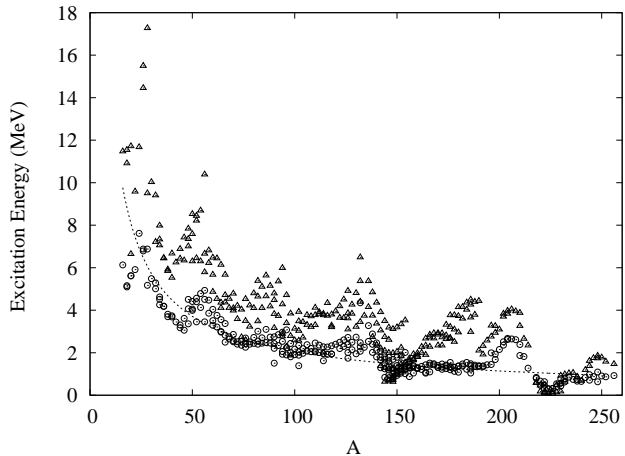}%
\caption{\label{E_vs_A} 
Octupole excitation energies as a function of mass number $A$.  Circles: 
experiment; triangles: theory.   
} 
\end{figure}

We now compare theory with the experimental data from the  
review by Kib\'edi and Spear~\cite{ki02}.  The excitation energies  
of the 284 tabulated nuclei with $Z\ge 8$ are shown in Fig. \ref{E_vs_A}, 
plotted as a function of $A$.  The data show a strong overall  
$A$-dependence as well as shell-related fluctuations.  The line 
shows a fit to the smooth trend in $A$ with the phenomenological 
parameterization $E(A) = 103/A^{0.85}$ MeV.  The most pronounced 
fluctuation about the trend is the rise and sudden drop near 
$A=208$; the drop is to low values is due to the extreme softness 
in the octupole mode. The theoretical energies, shown as triangles,  
replicate the overall trend with $A$ and the dramatic fluctuation at 
$A\sim208$.  However, overall the theoretical energies are too high, 
particularly in the light nuclei. 

\begin{figure}   
\includegraphics [width = 9cm]{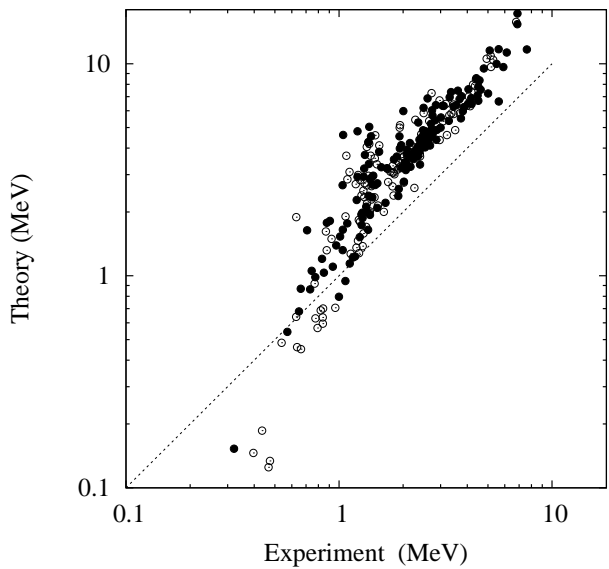}%
\caption{\label{e3} 
Octupole excitation energies, comparing the theory with experiment.  Filled 
circles are excitations with measured $B(E3)$ strengths; open circles are  
other identified octupole transitions\cite{ki02}. 
} 
\end{figure}

A more detailed comparison of theory and experiment may be seen on the 
scatter plot Fig.  \ref{e3}.  For excitation energies above 1 MeV, the 
theoretical values track the experimental but scaled by a factor.  Around 1 
MeV and below the theoretical values become closer to experiment.  The 
lowest energy measured excitations are in the Ra isotopes, where the 
theoretical HFB wave functions have static octupole deformations.  The 
theory reproduces the low energies to several hundred keV on an absolute 
energy scale, but does not do well on the logarithmic  energy 
scale shown in the figure. 
 
We also make some quantitative assessment of the performance of 
the theory, which should be useful in the future for comparing with 
other theories.  We use the same performance measures  
as was used to assess 
theories of quadrupole excitations\cite{sa07,de10}, namely to  
compare ratios of theoretical to experimental quantities on  
a logarithmic scale.  In terms of $R_E= \log(E(th)/E(exp))$ 
we determine the average value 
\be 
\label{rE} 
\bar R_E = \langle R_E \rangle 
\ee 
and the dispersion about the average, 
\be 
\label{sigmaE} 
\sigma_E = \langle (R_E - \bar R_E)^2\rangle^{1/2}.  
\ee  
The results are shown in  
Table \ref{re}.  The first line shows the comparison taking 
the full HW treatment on the theoretical side and the full 
data set on the experimental side.  One sees that the  
predicted energy is systematically too high, by a factor of 
$e^{0.44}\approx 1.6$.  This is similar to the situation with 
the quadrupole excitations.  There the understanding is that the 
wave function is missing components that would be included in  
collective theories using 
Thouless-Valatin inertial parameters.  There may be other reasons 
for the systematic overprediction here that we will come back to 
in Sect. \ref{discussion}.  The dispersion in the  
values is $\sigma_E\approx 0.4$, corresponding to errors in the 
ratio of theory to experiment of $-30\%$ to $+50 \%$.  This is larger 
than the global dispersion found for the GCM-based theories 
of quadrupole excitations. However, we saw in Fig. \ref{e3} that 
there are differences in the nuclear structure that are responsible 
for the variable performance of the theory.  Most importantly, 
the nuclei with calculated static octupole deformations should 
be treated separately.  Taking out these nuclei, the 
dispersion decreases dramatically, as shown on the second line 
of the Table.  A further distinction can be made between well-deformed and 
other nuclei, spherical and soft,  
respect to ordinary quadrupole deformations. 
A good theoretical indicator for deformed nuclei is the  
ratio of $4^+$ to $2^+$ excitation energies, called $R_{42}$. 
The values are available for the Gogny D1S interaction from the global  
study \cite{de10}, and we use them to set the condition $R_{42}>2.9$ 
to define the set of well-deformed nuclei.  The results are shown 
in the third and fourth rows of the table. One sees that the dispersion 
becomes even narrower for the nuclei in the nondeformed set.  Thus, we can 
claim that the HFB/GCM/HW methodology is quite successful for nondeformed 
nuclei, when allowing for the overall scale factor.  On the other hand, the 
deformed set is significantly poorer, with the average predicted energies 
higher and a larger dispersion.  A possible cause of this poorer performance 
could be the misidentification of transitions in deformed nuclei.  We have 
assumed here that all transitions are associated with the axially symmetric 
octupole operator ($K=0$).  As discussed in the next section, it is clear that some 
of the measured energies are for transitions with $K\neq 0$ (see also the 
$^{158}$Gd example).   
Since all the $K$ values in spherical nuclei 
are degenerate, this would explain the better overall agreement there.  
\begin{table}[htb] 
\begin{center} 
\begin{tabular}{|l|c|cc|cc|} 
\hline 
 & & \multicolumn{2}{c|}{HW} & \multicolumn{2}{c|}{MAP}\\ 
Selection & Number &  $\bar R_e $ &  $\sigma_e$& $\bar R_e $ &  $\sigma_e$ \\ 
\hline 
  all & 284  &  0.45  & 0.40 &&\\ 
 $\beta_3=0$  & 277  &  0.55  & 0.23 &0.59&0.22\\ 
 $\beta_3=0$, def. & 59  & 0.62 & 0.32 &0.75&0.26\\ 
 $\beta_3=0$, sph. & 196 & 0.52 & 0.19 &0.53 &0.17\\ 
\hline 
\end{tabular} 
\caption{\label{re} Performance of the HW theory for excitation energies 
compared to the experimental data tabulated in Ref. \cite{ki02}.   
The performance measures $r_E$ and $\sigma_E$ 
are given in Eq. (\ref{rE}) and (\ref{sigmaE}) of the text.  The performance  
of MAP is shown as well on lines 2-4 for subsets 
of nuclei selected by deformation criteria.  
} 
\end{center} 
\end{table} 

\subsection{Transition strengths} 
The octupole transition strength is computed from the proton octupole 
transition matrix element $ \langle o | \hat Q_3 \frac{1+t_z}{2}|e\rangle$. 
In a strongly deformed nucleus, the excitation is in a $K=0$ odd-parity  
band and the spectroscopic matrix element from the $3^-$ state in the 
band is given by 
\be 
\label{be3def} 
 B(E3, 3^-\rightarrow 0^+) = \frac{e^2}{4\pi}  
\langle o | \hat Q_3\frac{1+t_z}{2} |e\rangle^2. 
\ee 
This formula was used in Ref. \cite{eg91} to estimate the octupole 
transition strengths in Ra isotopes and other possible octupole-deformed 
nuclei.  On the other hand, if the state $|e\rangle$ is spherical, then 
the excitation induced by $Q_3$ gives a state $|o\rangle$ that has 
good angular momentum and the transition strength can be calculated directly as 
\be 
\label{be3sph} 
B(E3, 3^-\rightarrow 0^+) = \frac{7 e^2}{4\pi} \langle o | \hat Q_3 
\frac{1+t_z}{2}|e\rangle^2. 
\ee 
Notice that this is a factor of 7 larger than Eq. (\ref{be3def}). 
The reason for the difference is that Eq. (\ref{be3sph}) gives 
a total octupole transition strength, while Eq.(\ref{be3def}) only 
gives the transition strength for the $K=0$ components.   
 
\begin{figure} 
\includegraphics [width = 9cm]{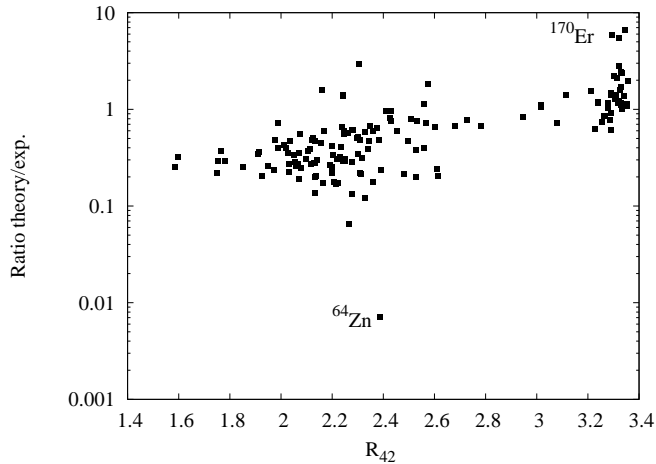} 
\caption{\label{be3} 
Ratio of theoretical octupole transition strength to experimental,  
with the theoretical strength obtained using Eq.~(\ref{be3def}).  The 
horizontal axis is the ratio $R_{42}$ from the theory of Ref. \cite{de10}. 
Experimental $B(E3)$ values are from Ref. \cite{ki02}. 
} 
\end{figure}

Besides these limiting cases, there are soft nuclei which should 
fall in between.  Thus, it is imperative to restore good angular 
for the theory to have a global applicability.  While angular  
momentum projection has been carried out in the past\cite{eg93,su94,eg96}, 
it is beyond the scope of this article.  Instead, we examine here 
the range of predicted values using a theoretical marker of the deformation 
to distinguish nuclei falling in the different categories.  Fig.  
\ref{be3} shows the ratios of theoretical to experimental 
$B(E3)$ values, using the experimental data set from Ref. \cite{ki02} 
and Eq. (\ref{be3def}) for the theory.  The  
data is plotted as a function of the quantity $R_{42}$, the ratio of the 
lowest $4^+$ to $2^+$ excitation energies.    
Values around 2 or less are characteristic 
of spherical nuclei, while strongly deformed nuclei have $R_{42} \ge 3$. 
We take the values for $R_{42}$ from the spectroscopic calculations of  
Ref. \cite{de10}, based on HFB/GCM with the same Gogny D1S interaction 
used for the theory here. 
The plot show a lot of scatter, but one can see two groups of nuclei, 
the lefthand representing deformed nuclei.  There is a trend visible 
in the $B(E3)$ ratios consistent with the above discussion.

To make the analysis more quantitative, we examine the logarithmic 
averages $\bar R$ dividing the nuclei into two group according to 
$R_{42}$.  The results are shown in Table \ref{be3_perf}.  Since we 
use Eq. (\ref{be3def}) to determine $R$, we should find $\bar R = 0$ 
for the first row of the Table.  In fact, the average is about 40 \% 
high.  For the second row, if all the nuclei were spherical, the  
strength should be a factor of 7 larger.  This implies that the 
$\bar R$ calculated with the deformed formula should give a  
value $0.33-\log(7)= -1.6$.  The value found, -0.99, shows that there 
is an important effect of the deformation but that it to simplistic 
to assume that these nuclei are all spherical.   
 
\begin{table}[htb] 
\begin{center} 
\begin{tabular}{|c|c|cc|} 
\hline 
Selection & Number &  $\bar R$ &  $\sigma $ \\ 
\hline 
Deformed, $R_{42}> 2.9$ & 41 &  0.34   &  0.5 \\ 
Other, $R_{42}< 2.9$ & 112 &  -0.99   &  0.7 \\ 
\hline 
\end{tabular} 
\caption{\label{be3_perf} Ratio of theoretical to experimental $B(E3)$ 
strengths. The second column is the number of nuclei in the data set. 
} 
\end{center} 
\end{table} 
 
We note that the enhancement of the $B(E3)$ for the less deformed nuclei 
is evident in the projected calculations for $^{16}$O (\cite{eg93}) and 
Pb isotopes near $A=208$ (\cite{eg96}).  Also, in Ref. \cite{sk93a} the 
authors remark on a strong disagreement between theory and experiment 
for $^{96}$Zr.  This is the case if one uses Eq. (\ref{be3def}), but  
that nucleus is spherical according to the $R_{42}$ criterion and Eq. 
(\ref{be3sph}) gives a satisfactory agreement. 
agreement is satisfying  
 
It is of interest to examine the nuclei that deviate most strongly from  
the theory.  In Fig. \ref{be3} there is a group of three outlier nuclei 
in the upper right-hand corner.  The nuclei are $^{170}$Er and its 
neighbors.   In these cases, the experimental transitions are  
likely to be to excited states with $K\neq0$.  The lowest $1^-$  
excitation in $^{170}$Er at 1.26 MeV has a $K=1^-$ character, and 
the first $K=0^-$ is higher by 0.6 MeV.  There are some studies 
in the literature in which the $K$-dependence of the octupole excitation 
is examined\cite{sk93b,ya01,zb06}.  In Ref. \cite{ya01,zb06} the $K=0^-$ 
bands were found to be higher in energy than other $K$ values. 
 
The other glaring anomaly is the nucleus $^{64}$Zn at  
$R_{42}\approx 2.4$, which has a grossly underpredicted $B(E_3)$. 
It turns out that the quadrupole deformation of this nucleus changes sign  
as $\beta_3$ is increased.  The ground 
state at $\beta_3=0$ is oblate, but it switches to another minimum 
with a prolate shape at moderate values of $\beta_3$.  The very small 
predicted $B(E3)$ is due to the small overlap between the oblate and 
prolate configurations.  Clearly, the GCM must include explicitly  
both quadrupole and octupole degrees of freedom to properly treat 
this nucleus. A few other nuclei with similar Z values show the same 
behavior. We note that the $B(E3)$ comes out much closer to  
experiment if both even and odd states are taken from configurations 
having the same sign of quadrupole moment. 
\section{Discussion} 
\label{discussion} 
 
We have demonstrated that a global theory of the octupole degree of 
freedom can be constructed using the HFB/GCM/HW methodology.  The 
theory reproduces the secular trend of the excitations, the effects 
of an incipient static octupole deformation, and the most visible shell  
effects.  However, the theory has obvious deficiencies.  Most notably, 
we require a overall scaling factor of 1.6 to make quantitative comparison 
with experiment.  It is urgent to understand what physics is needed to 
make predictions on an absolute energy scale.  There are several possible  
reasons for the absolute errors.  One is the Hamiltonian itself.  Besides 
the Gogny interaction, there have been calculations with the BCP interaction,  
interactions from 
the Skyrme family and from relativistic mean-field theory.  Ref. 
\cite{ro10b} found that the D1S Gogny interaction and the BCP interaction 
gave significant differences in the odd-parity excitations of Ra isotopes. 
The 
Gogny interaction is guided by nuclear Hartree-Fock theory, and one of 
the characteristics is a nucleon effective mass less than the physical 
mass.  This implies that single-particle 
excitation energies will be higher than for a non-interacting system, 
and these effects could carry over to the collective excitations as 
well.  We note that 
the calculation of the $^{208}$Pb in Ref. \cite{he01} using a Skyrme  
interaction with a similar effective mass to D1S agrees with our results.  
However, the Relativistic Mean Field Hamiltonian also has a 
small effective mass, but excellent agreement was obtained for $E_3$ in  
an isotone chain by (Q)RPA \cite{an09}.  
 
This brings up another source of systematic error in the GCM/HW, 
the restriction of the degrees of freedom in the excitation to a single 
variable. It is well-known in the theory of quadrupole excitations 
that time-odd components must be included in the wave function to 
obtain good moments of inertia \cite{ro00}.  For large amplitude deformations,  
this can be achieved by self-consistent cranking.  
When no time-odd components are allowed in the angular  
momentum projected (AMP) GCM calculation the excitation energy is  
stretched with respect to standard cranking calculations by a factor of  
around 1.4. This correction factor is compatible with the  
discrepancies observed between our results and the experiment in the  
case of $^{158}$Gd as well as with the overall 1.6 factor for the 
negative parity excitation energies discussed previously. 
 
More generally, one can introduce methods that would reduce to  
(Q)RPA in the small amplitude limit.  The raises the question of how  
well (Q)RPA would perform in a global context.  As shown in the  
Appendix, for a large fraction of nuclei the GCM/HW methodology is  
essentially equivalent to (Q)RPA in a single collective variable.   
For these nuclei, the (Q)RPA is justified and is very likely to give  
lower excitation energies.     
 
The interaction of the octupole with the quadrupole degree of freedom 
is an interesting problem that appears in several contexts in our 
study.  First, the HFB static quadrupole deformation of many nuclei 
invalidates a spectroscopic interpretation of the observables for 
the physical angular momentum eigenstates of the system.  We saw 
this most directly in the discussion of the $B(E3)$ transition 
strengths.  The solution is to carry out angular momentum projection. 
Another aspect missing from our study is 
the inclusion of $K\neq0$ excitations in deformed nuclei.  This has been  
done in HFB-BCS in Ref. \cite{sk93b,da99,zb06} and in HFB in Ref. \cite{ya01}. 
Since $K\neq0$ bands can fall below the $K=0$ octupole excitation band, 
it is essential for a complete theory of the octupole excitations in 
deformed nuclei.   
 
Some aspects of the quadrupole-octupole mixing may require a two-dimensional 
GCM to describe properly.  It was clear in the light nuclei that 
octupole and quadrupole deformations are strongly coupled in forming 
alpha-clusters.  Also we found that the severe problem describing the 
$B(E3)$ in $^{64}$Zn could be traced to the coupling.  We note that 
the two-dimensional GCM has been implemented in the past.  In Refs.  
\cite{me95} the coupled GCM was applied to the complex spectroscopy of 
the nucleus $^{194}$Pb.  Also, the microscopic theory of asymmetric 
fission\cite{go05} requires at least a two-dimensional GCM. 
 
One last aspect of the theory should be mentioned. We have seen in the  
examples that the correlation energy of the ground state associated with 
the $K=0$ octupole excitation is of the order 
of one MeV.  This can have an important influence on the 
theory of the nuclear masses.  We plan to investigate the systematics 
of the correlation energy in a future publication. 
 
The approximation of a single degree of freedom can break down in different ways.   
One 
is if the coupling between different multipoles is important in determining 
the configurations.  This is the case for light alpha-particle nuclei. 
Fig.~\ref{b23} shows the $\beta_2$ and $\beta_3$ deformations of the  
GCM configurations for $^{16}$O and $^{20}$Ne.   
It may be seen that the 
$\beta_3$ deformation carries a $\beta_2$ deformation along with it 
for all but the smallest values of $\beta_3$.  Whether this is physical 
or not depends on the matrix elements of the interaction connecting  
the different configurations.  For $^{16}$O, the admixtures are 
perturbative and thus should not change $\beta_2$.  However, if the 
configurations are sampled on coarse mesh, there will be significant 
admixture of quadrupole excitations and the Gaussian overlap approximation 
will fail.  As mentioned in the Introduction, a two-dimensional treatment 
of the GCM taking the quadrupole and octupole deformations as independent 
variables is important in fission\cite{go05}.  It has also been carried 
out for the $^{194}$Pb nucleus\cite{me95} which is very soft with respect 
to quadrupole deformations. 
 
The single-operator approximation is also problematic due to the  
fragmentation of octupole strength in the full spectrum.  Roughly 
speaking, the octupole strength has two important branches: the low 
collective excitation that is under study here, and the high-lying 
excitation characterized as $3\hbar \omega$ in the harmonic oscillator 
model.  Our generating field introduces amplitudes of both into the 
constrained wave function.

\section{Acknowledgments} 
We thank H. Goutte, P.-H. Heenen and W. Nazarewicz for reading and  
comments on the manuscript.   
This work was supported in part by the U.S. Department of Energy under Grant 
DE-FG02-00ER41132, and by the National Science 
Foundation under Grant PHY-0835543. 
The work of LMR was supported by MICINN (Spain) under  
grants Nos. FPA2009-08958, and FIS2009-07277, as well as by  
Consolider-Ingenio 2010 Programs CPAN CSD2007-00042 and MULTIDARK  
CSD2009-00064.

\section*{Appendix: Simplified approximations and limits} 
 
It is important to understand the limiting behavior of any computationally 
demanding theory, both to check the reliability of the calculations as 
well as to see whether approximations are justified that would simplify 
the calculations.  For the GCM/HW methodology, the theory becomes analytic 
or nearly so if a few conditions are met.  One requirement  is that 
there be only a single degree of freedom necessary to describe the 
excitation of the system.  There are simple Hamiltonians that satisfy 
this condition. Examples are the Lipkin model\cite{RS}, \cite{ro92}, where the 
degree of freedom is the number of particles in the excited orbital, 
and the two-particle problem treated in Ref. \cite{ha00} where the 
degree of freedom is the center-of-mass displacement.  In the last 
model and other like it the theory becomes analytic and reduces to  
RPA the if overlap integrals satisfy 
the Gaussian Overlap Approximation and the matrix elements of the  
Hamiltonian reduce to a quadratic functions times the overlap.   
In fact the relation to RPA remains even if there are many degrees of 
freedom in the GCM \cite{br68,ja64}.   
 
To make the discussion concrete, let us assume that there is a single 
continuous degree of freedom $q$ and we can write the overlap integral 
and the Hamiltonian matrix element as 
\be 
\label{goa} 
\langle q' | q \rangle = e^{-(q-q')^2/q_s^2} 
\ee 
\be 
\label{simple-h} 
\frac{\langle q' | H |q \rangle}{\langle q' | q  \rangle}= 
E_0 + \frac{1}{2} v (q + q')^2- \frac{1}{2} w (q - q')^2 
\ee  
The solution obtained by the Hill-Wheeler construction is identical to 
the solution of the RPA equation for the operator $\hat Q$ that generates 
the GCM states $|q\rangle$.  The HW wave functions have the form of 
Gaussians in the variable $q$ and the excitation energy is given by 
\be 
\label{rpa} 
\hbar\omega_{RPA} = q_s^2 \sqrt{v w}. 
\ee 
Let us now compare with the MAP approximation.  Here one first calculates  
projected energies  
as a function of 
$q$, 
\be 
\label{map-e} 
\frac{\langle e | H | e \rangle}{\langle e |   e \rangle} 
=2 v q^2 \frac{v-w e^{-4(q/q_s)^2}}{1+e^{-4(q/q_s)^2}} 
\ee 
and  
\be 
\frac{\langle o | H | o \rangle}{\langle o |   o \rangle} 
=2 v q^2 \frac{v+w e^{-4(q/q_s)^2}}{1-e^{-4(q/q_s)^2}} 
\ee 
The energies are then minimized with respect to $q$.  The results for a 
range of values of the ratio $w/v$ are given in Table \ref{MAP}.  The 
ratios $q_0/q_e$ are close to $\sqrt{3}$, which may reflect the  
harmonic oscillator character of the exact HW wave functions.  In the 
last columns we compare the MAP excitations energy with the RPA values. 
They are remarkably close.   
 
\begin{table}[htb] 
\begin{center} 
\begin{tabular}{|c|cc|cc|cc|} 
\hline 
$w/v$ & $q_e$ &$E_e$ & $q_o$&$E_o$ 
& $E_o-E_e$ & $\hbar\omega_{RPA}$ \\ 
\hline 
1.5 &  0.226 & -0.0125 & 0.390  & 1.212 & 1.225 & 1.225  \\ 
2.  &  0.292 & -0.0421 & 0.509  & 1.373 & 1.415 & 1.414   \\ 
4.  &  0.400 & -0.232  & 0.716  & 1.782 & 2.01 & 2.00  \\ 
8.  &  0.469 & -0.721  & 0.870  & 2.207 & 2.93 & 2.83   \\ 
\hline 
\end{tabular} 
\caption{\label{MAP} The MAP solution in the harmonic limit. 
Deformations are in units of $q_s$ and energies are in units of 
$vq_s^2$.  The last column shows the (Q)RPA excitation energy,  
Eq (\ref{rpa}). 
} 
\end{center} 
\end{table} 
As a general conclusion, we find that if the MAP conditions are satisfied, 
the energies are close to the RPA performed with a single collective 
variable.  For those nuclei, it would better to extend the space  
for the calculation using more RPA degrees of freedom than by 
going to large amplitudes in a single collective variable.    
 
\begin{figure} 
\includegraphics [width = 9cm]{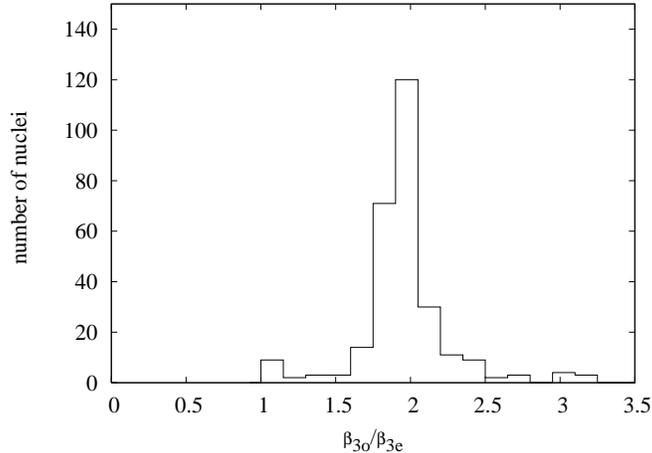} 
\caption{\label{b3oe} 
Ratio of MAP deformations $\beta_{3p}/\beta_{3m}$ for nuclei with  
measured $E_3$ \cite{ki02}. 
} 
\end{figure}

It would be nice to find a criterion to test  
for validity of the simplified treatment.  The first condition we can 
check is the ratio $q_o/q_e$.  This is graphed in Fig. \ref{b3oe} for 
the 284 nuclei tabulated in Ref. \cite{ki02}. 
There is a strong peak at $\beta_{3m}/\beta_{3p}\approx 1.9$.  This is 
slightly higher than the single-mode (Q)RPA, but still close enough 
to make a further investigation of the quadratic Hamiltonian approximation. 
There are also wings on the distribution extending from 0.9 ($^{16}$O) to 
3.2 ($^{230}$U).  Excluding the wings below 1.7 and above 2.2, the peak 
contains 80 \% of the measured nuclei. 
 
\begin{figure} 
\includegraphics [width = 9cm]{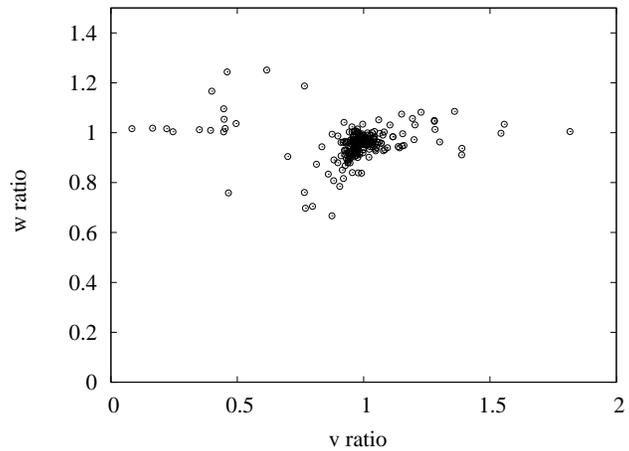} 
\caption{\label{quadratic} 
Ratio of MAP deformations $\beta_{3m}/\beta_{3p}$ for nuclei with  
measured $E_3$ \cite{ki02}. 
} 
\end{figure}

To examine the validity of the quadratic approximation, we  
compared the extracted coefficients $v q_s^2$ and 
$w q_s^2$ at the two deformations $\beta_{3p}$ and $\beta_{3m}$.  If 
the quadratic approximation is valid, they should be equal.  For example, 
the values of $\beta_{3p}$ and $\beta_{3m}$ at the closest mesh points 
are 0.0375 and 0.075, respectively.  The values of $v\beta_{3p}^2$ and 
$w\beta_{3p}^2$ extracted at that mesh point are 0.23 MeV and 1.72 MeV, 
respectively.  The corresponding numbers for $\beta_{3m}$ are 
0.94 MeV and 7.20 MeV, very close to 4 times the values at $\beta_{3p}$. 
This is just what is expected given $\beta_{3m}/\beta_{3p}= 2$, showing 
that $^{208}$Pb satisfies the conditions for the quadratic Hamiltonian. 
With these values for $v$ and $w$, the RPA energy formula Eq. (\ref{rpa}) gives 
3.9 MeV, close to the GCM/HW value of $4.0$ MeV.  The results for the 
nuclei within the peak of Fig. \ref{quadratic} is shown as a scatter plot of the 
ratios.  In general, the $w$ term follows a quadratic dependence very well. 
The $v$ term can have large deviations, particularly for nuclei that are 
soft to octupole deformations.  However, for most of the nuclei, the quadratic approximation is valid 
to an accuracy far better than needed, given the overall performance of  
the theory in non-octupole deformed nuclei at the 25\% level in the scaled  
energies.   


\begin{thebibliography}{99} 
\bibitem{ma83} S.~Marcos, H.~Flocard, and P.H.~Heenen, Nucl. Phys. {\bf 
A410} 125 (1983). 
\bibitem{he01} P.-H.~Heenen, et al., Eur. Phys. J. A {\bf 11} 393 (2001). 
\bibitem{eg91} J.L.~Egido and L.M.~Robledo, Nucl. Phys. {\bf A524} 65 
(1991). 
\bibitem{sk93b} J. Skalski, et al., Nucl. Phys. {\bf A551} 109 (1993). 
\bibitem{he94} P.-H.~Heenen, et al., Phys. Rev. C {\bf 50} 802 (1994). 
\bibitem{ga98} E. Garrote, J.L.~Egido, and L.M.~Robledo, Phys. Rev. Lett. 
{\bf 80} 4398 (1998). 
\bibitem{se02} A.P~Severyukhin, C.~Stoyanov, V.V.~Voronov, and N. Van 
Giai, Phys. Rev. C{\bf 66} 034304 (2002). 
\bibitem{co03} G. Col\`o, et al., Nucl. Phys. {\bf A722} 111c (2003). 
\bibitem{an09} A.~Ansari and P.~Ring, Mod. Phys. Lett. A {\bf 24} 3103 
(2009). 
\bibitem{bu96} P.A.~Butler and W.~Nazarewicz, Rev. Mod. Phys. {\bf 68} 
349 (1996). 
\bibitem{sa07} B. Sabbey, M. Bender, G.F. Bertsch, and P.-H. Heenen, 
Phys. Rev. C 75, 044305 (2007).  
\bibitem{de10} J.-P.~Delaroche, et al., Phys. Rev. C {\bf 81} 014303 (2010). 
\bibitem{go05}  H.~Goutte, J.F.~Berger, P.~Casoli, and D.~Gogny,  
Phys. Rev. C {\bf 71} 024316 (2005). 
\bibitem{me95} J. Meyer, et al., Nucl. Phys. {\bf A588} 597 (1995). 
\bibitem{pe08} S. P\'eru and H. Goutte, Phys. Rev {\bf C 77}, 044313 (2008) 
\bibitem{ro02b} R. Rodr\'\i guez-Guzm\'an, J.L. Egido and L.M. Robledo, 
Nucl. Phys. {\bf A709}, 201 (2002) 
\bibitem{ro10b} L.M.~Robledo, et al., Phys. Rev. C {\bf 81} 034315 (2010). 
\bibitem{onishi}In the present application, 
there is no phase ambiguity in applying the Onishi formula  
because the configurations are time-reversal invariant. 
\bibitem{balian} R.~Balian and E.~Brezin, Nuovo Cimento B{\bf64} 37 (1969). 
\bibitem{bo90} P. Bonche, J. Dobaczewski, H. Flocard, P. -H. Heenen, and J. Meyer, 
               Nucl. Phys. {\bf A 510}, 466 (1990)                 
\bibitem{ro07} L.M. Robledo,  
               Intl. J. of Mod. Phys. {\bf E 16}, 337 (2007). 
\bibitem{ro10a} L.M.~Robledo, J. Phys. G {\bf 37} 064020 (2010). 
\bibitem{epaps} If accepted by Phy. Rev. C, the files will be accessible 
in the supplementary data accompanying the article. 
\bibitem{wu} 
The transition strength in Weisskopf units $W(E3)$ is given by the  
formula $  W(E3) = 2.404\times 10^6 B(E3,\uparrow)/A^2$. 
\bibitem{ki02} T. Kib\'edi and R. Spear, Atomic Data and Nuclear Data 
Tables {\bf 80} 35 (2002); R.H.~Spear, Atomic Data and Nuclear Data 
Tables {\bf 42} 55 (1989). 
\bibitem{mo08} P. M\"oller, R. Bengtsson, et al., At. Data Nucl. Data Tables 
{\bf 94} 758 (2008). 
\bibitem{zh10} W.~Zhang, Z.~Li, Q.~Zhang, and J.~Meng, Phys. Rev. C {\bf 81} 
034302 (2010). 
\bibitem{eg93} J.L.~Egido, L.M.~Robledo, and Y.~Sun, Nucl. Phys. {\bf A560} 
253 (1993). 
\bibitem{su94} Y.~Sun, L.M.~Robledo, and J.L.~Egido, Nucl. Phys. {\bf A570} 
305c (1994). 
\bibitem{eg96} J.L.~Egido, V.~Martin, L.M.~Robledo, and Y.~Sun, Phys. Rev. C 
{\bf 53} 2855 (1996). 
\bibitem{sk93a} J.~Skalski, P.-H.~Heenen, P.~Bonche, Nucl. Phys. {\bf A559} 
221 (1993). 
\bibitem{zb06} K.~Zberecki, P.~Magierski, P.-H.~Heenen, and N. Schunck, 
Phys. Rev. C {\bf 74} 051302 (2006). 
\bibitem{ya01} M.~Yamagami,K~Matsuyanagi,M.Matsuo, Nucl. Phys. {\bf A693} 
579 (2001). 
\bibitem{ro00} R. Rodr\'\i guez-Guzm\'an, J.L. Egido and L.M. Robledo, 
Phys. Rev.  {\bf C62}, 054319 (2000) 
\bibitem{da99} H.~Dancer et al., Nucl. Phys. {\bf A654} 655 (1999). 
\bibitem{RS} P. Ring and P. Schuck, "The Nuclear Many-Body Problem",  
(Springer, NY), 1980, Sec. 10.7.5. 
\bibitem{ro92} L.M.~Robledo, Phys. Rev. C{\bf 46} 238 (1992). 
\bibitem{ha00} K.~Hagino and G.F.~Bertsch, Phys. Rev. C {\bf61} 
024307 (2000). 
\bibitem{br68}  
D.M.~Brink and A. Weiguny, Nucl. Phys. {\bf A120} 59 (1968). 
\bibitem{ja64} Another more restricted derivation is given by 
B. Jancovici and D.H. Schiff, Nucl. Phys. {\bf 58} 678 (1964). 
\end{thebibliography}
\end{document}